\documentclass[twocolumn,printnumbers,amsmath,amssymb,showpacs]{revtex4}

\newcommand{\be}{\begin{equation}}
\newcommand{\ee}{\end{equation}}
\newcommand{\ba}{\begin{eqnarray}}
\newcommand{\ea}{\end{eqnarray}}
\usepackage{graphicx}
\usepackage{dcolumn}
\usepackage{bm}
\begin{document}

\title{Quantum buckling}


\author{N. Upadhyaya and V. Vitelli} 
\affiliation{Instituut-Lorentz, Universiteit Leiden, Postbus 9506, 2300 RA Leiden, The Netherlands.}

\begin{abstract} 
\noindent We study the mechanical buckling of a two dimensional membrane coated with a thin layer of superfluid. It is seen that a singularity (vortex or anti-vortex defect) in the phase of the quantum order parameter, distorts the membrane metric into a negative conical singularity surface, irrespective of the defect sign. The defect-curvature coupling and the observed instability is in striking contrast with classical elasticity where, the in-plane strain induced by positive (negative) disclinations is screened by a corresponding positive (negative) conical singularity surface. Defining a dimensionless ratio between superfluid stiffness and membrane bending modulus, we derive conditions under which the quantum buckling instability occurs. An ansatz for the resulting shape of the buckled membrane is analytically and numerically confirmed. 
\end{abstract}

\maketitle

\section{Introduction}

\noindent The rapid trend towards the miniaturization of electro-mechanical systems has spurred a flurry of theoretical activity aimed at investigating quantum mechanical effects in the context of more classical subjects like heat transfer and mechanical stability \cite{Blencowe:2004p13,Kim:2008}. Common building blocks for these devices are carbon based materials, such as nanotubes and graphene, essentially two-dimensional elastic sheets which are often bent or wrinkled \cite{Meyer:2007p30,deJuan:2007p33}. Much effort has been directed towards understanding how quantum surface states are affected by the underlying curvature of these spaces \cite{DaCosta:1981p29,Neto:2009p32,Turner:2010p17,Vitelli:2004p20,Guinea:2010}. However, less attention has been devoted to the converse problem: can quantum mechanical effects modify the shape and mechanical stability of nano-structures?
\\
\\
\noindent These questions have so far been relegated to the fringe of mainstream engineering applications, since it is challenging to probe experimentally the regime where the characteristic energy of quantum effects is comparable to the bending energy. Furthermore, the interplay between a quantum order parameter and geometry is a more subtle theoretical problem than its classical counterpart.  Quantum mechanical degrees of freedom {\it live} in an internal space distinct from the local tangent plane of the underlying substrates. As a result, coupling mechanisms between the in-plane quantum order parameter and curvature are typically less intuitive. By contrast, classical buckling is the paradigmatic example of how elastic stresses in a crystal or liquid-crystal monolayer are {\it screened} by curvature \cite{Klein:2007p45,Witten:2007p25,Frank:2008p28,Vitelli:2006p21,Santangelo:2007p22,Bausch:2003p10,Bowick:2009p27,Irvine:2010p34,DeVries:2007p26}.
\\
\\
\noindent In this letter, we demonstrate a quantum analogue of buckling and trace its distinctive physical and mathematical origins. Specifically, we consider the two dimensional (2D) order parameter, $\psi(\mathbf{\rho})=|\psi| e^{i\theta(\mathbf{\rho})}$ that describes superfluid or superconducting phases of a quantum condensate and show that, the presence of isolated point defects (vortices or anti-vortices) where the amplitude of the quantum order parameter vanishes,  causes the substrate to buckle, apparently resembling the buckling of flexible membranes by classical crystalline defects. However, there is a crucial qualitative difference between the two phenomena (summarized pictorially in Fig. 1) that stems from the distinct physical mechanisms responsible for buckling. Buckling induced by defects screening requires that, positive (negative) disclinations (eg. 5 or 7-fold coordinated atoms in a crystalline monolayer) induce buckling into conical singularities of positive (negative) Gaussian curvatures \cite{Park:1996p18,Deem:1996p15,Seung:1988}: screening requires that the sign of the defect charge is matched by the sign of the curvature (see Fig. 1 a-b). In contrast, our stability analysis reveals that despite the lack of an explicit defect screening mechanism, quantum defects (whose flow lines are shown in Fig. 1c and d respectively) will induce buckling into a conical singularity of negative Gaussian curvature, independently of the sign of the defect charge. 
\\
\\
\noindent Consider for simplicity, a 2D flexible liquid membrane of vanishing surface tension supporting a quasi two-dimensional quantum condensate. The total energy functional $H$ can be written as the sum of contributions from the membrane bending energy $H_{\kappa}$ and the condensate kinetic energy $H_{v}$, respectively given by
\begin{eqnarray}
H_{\kappa} &=& \frac{\kappa}{2}\int d^2\mathbf{u} \sqrt{g} \ M^2(\mathbf{u}) , \label{Hamiltonian1} \\
H_v &=& \frac{K}{2}\int d^2\mathbf{u} \!\!\!\! \quad \sqrt{g} \!\!\!\! \quad g^{\alpha \beta} \!\!\!\! \quad \partial_{\alpha}\theta(\mathbf{u}) \!\!\!\! \quad \partial_{\beta} \theta(\mathbf{u}) \!\!\!\! \quad . \label{Hamiltonian2}
\end{eqnarray}
Here, $\mathbf{u}=\{u^1,u^2\}$ is a set of two dimensional coordinates specifying positions $\mathbf{R}(\mathbf{u})$ in the plane of the membrane, $g_{\alpha \beta}=\partial _{\alpha}\mathbf{R}\cdot\partial _{\beta}\mathbf{R}$ and $g=\text{det}(g_{\alpha \beta})$ denote the metric tensor and its determinant, $M(\mathbf{u})$ is the extrinsic mean curvature \cite{Kamien:2002}, $\kappa$ is the membrane bending rigidity, $K=\frac{\rho_s\hbar}{2m^2}$ is the superfluid stiffness constant given in terms of the atomic mass $m$ of the superfluid and density $\rho_s$, and $\partial_{\alpha}\theta(\mathbf{u})$ gives the local superfluid velocity. \\
\\
\noindent In order to grasp intuitively the origin of the quantum buckling instability, consider Eq. (\ref{Hamiltonian2}) for the case of an isolated vortex of topological charge $q=\pm 1$ at the tip of a conical singularity: an azimuthally symmetric surface denoted by a height function (out of plane shift) $h(\rho,\phi)=m\rho$, where we have used 2D polar coordinates $\mathbf{u}=\{\rho,\phi\}$. Note that, the Gaussian curvature for this surface is a delta function that vanishes everywhere except at the tip of the cone. Due to the azimuthal symmetry, we expect the elastic variable $\theta=q\phi$ to retain its flat space form, so that $\nabla\theta= \frac{q}{\rho} \hat{\mathbf{e}}_{\phi}$, where $\hat{\mathbf{e}}_{\phi}$ is the angular unit vector in polar coordinates. With the metric expressed in terms of the slant length of the cone $l=\sqrt{1+m^2}\rho$, we can evaluate Eq. (\ref{Hamiltonian2}) to obtain
\begin{eqnarray}
E_v &=& \pi Kq^2\sqrt{1+m^2}\ln\left(\frac{R}{a_0}\right) \!\!\!\! \quad ,
\label{cone1}
\end{eqnarray}
where, $a_0$ is a microscopic cut-off length (of the order of the vortex core radius) and $R>>a_0$, is the size of the membrane. We see that the energy required for the vortex (anti-vortex) to occupy the tip of a conical singularity is always greater than its flat-space counterpart by a positive definite factor $\sqrt{1+m^2}$ in Eq. (\ref{cone1}) \footnote{The comparison is meaningful in our assumed limit $R\rightarrow\infty$ where one does not need to worry about measuring the size of the cone along itself or on its base.}. This simple calculation demonstrates that it is not energetically favourable for the vortex to buckle the membrane into a surface with a positive delta Gaussian curvature (the positive definite bending energy in Eq. (\ref{Hamiltonian1}), only adds an extra penalty). 
\\
\\
\noindent Contrast the result obtained in Eq. (\ref{cone1}) with the corresponding one for a liquid crystal membrane. In the liquid crystal case, the order parameter $\theta$ describes the orientation of a vector, not the (scalar) phase of a wave-function. This distinction implies that the elastic variable $\theta$ must explicitly couple with the underlying curvature and thus, each instance of $\partial _{\alpha}\theta$ in Eq. (\ref{Hamiltonian2}) appears in the form $\partial_{\alpha} \theta-A_{\alpha}$, where the connection $A_{\alpha}$ is a geometric gauge field whose curl equals the Gaussian curvature. As a result of this difference, the leading order correction in $m$, appears with a minus sign in Eq. (\ref{cone1}) and it can be sufficiently large to overcome the bending energy cost \cite{Park:1996p18,Deem:1996p15}. This is the mathematical mechanism responsible for the classical buckling of liquid crystal as well as crystalline membranes: the geometric gauge field couples elastic defects to Gaussian curvature via cross terms in $(\partial_{\alpha} \theta-A_{\alpha})^2$, thereby providing a direct mechanism for screening the defect charge. 
\begin{figure}
\begin{center}
\includegraphics[width=90mm]{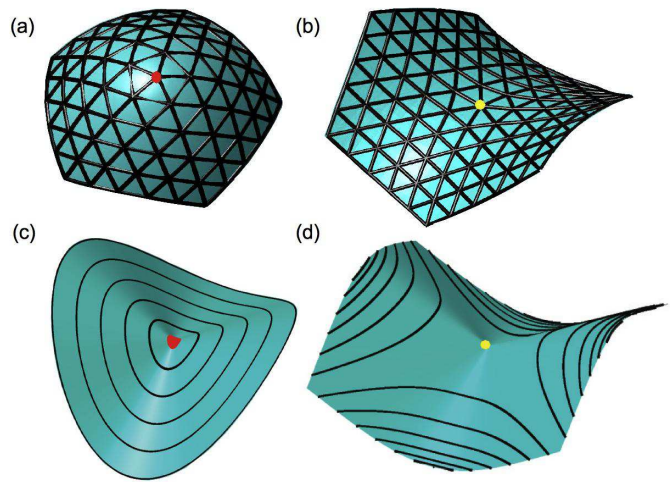}
\caption{\label{fig:fig1} (Color online) Illustrative plots for membrane buckling corresponding to : (a) positive disclination,(b) negative disclination, (c)vortex and (d)anti-vortex. Positive (negative) topological defects are shown with red (yellow) dots, where a positive (negative)  disclination is screened by a surface with positive (negative) Gaussian curvature in the top row, while a vortex (anti-vortex) buckles the underlying membrane into a negative Gaussian curvature surface in the bottom row.}
\end{center}
\end{figure}  
\\
\\
Can there be buckling in the absence of a geometric gauge field? In order to answer this question, we take a more systematic and versatile approach to track the curvature correction estimated in Eq. (\ref{cone1}). For general surfaces and defects, curvature corrections arise from a conformal anomaly obtained by puncturing the domain of integration around defect cores in Eq. (\ref{Hamiltonian2}), such that the energy remains finite. As a result,  the vortex free energy becomes the sum of two contributions, $E_v = E_f + E_s$, where $E_f$ is the flat space energy of the vortex and, $E_s = -Kq^2V(\mathbf{u})$ is a self-energy contribution expressed in terms of the geometric potential $V(\mathbf{u})$ that satisfies a covariant form of Poisson's equation:
\begin{eqnarray}
D_{\alpha}D^{\alpha}V(\mathbf{u})=G(\mathbf{u}) \!\!\!\! \quad.
\label{poisson}
\end{eqnarray}
Here, the negative of the Gaussian curvature $G(\mathbf{u})$ plays a role analogous to the electrostatic charge density \cite{Vitelli:2004p20,Turner:2010p17}. 
\\
\\
\noindent We now solve Eq. (\ref{poisson}) for a family of surfaces whose height function is described by $h(\rho,\phi)=\rho f(\phi)$. The geometric potential for a singular distribution of Gaussian curvature reads 
\begin{eqnarray}
V(\rho) = -2\pi s \Gamma(\rho + a_0,\rho) \!\!\!\! \quad,
\end{eqnarray}
where, the integrated Gaussian curvature $s$ can be obtained using the Gauss-Bonnet theorem \cite{Park:1996p18},
\begin{eqnarray}
s = 1 - \frac{1}{2\pi}\int^{2\pi}_0\ d\phi\frac{\left[1+f^{'2}(\phi) - f(\phi)f''(\phi)\right]}{\left[1+f^{'2}(\phi) + f^2(\phi)\right]^{1/2}\left[1+f^{'2}(\phi)\right]} \!\!\!\! \quad, \nonumber
\label{spm}
\end{eqnarray}
and the Green's function $\Gamma$ evaluated at the core of the defect takes the form $\Gamma(\rho+ a_0, \rho)=\lim_{\rho\to a_{0}}-A\left(\ln\frac{\rho}{a_0}-\ln\frac{R}{a_0}\right)= A\ln\frac{R}{a_0}$ \footnote{The $\phi$ dependence possible for non-axisymmetric surfaces drops out due to the divergent logarithmic term \cite{Park:1996p18,Deem:1996p15}}. The coefficient $A$ is then given by \cite{Park:1996p18}
\begin{eqnarray}
A = \left[\int^{2\pi}_0\ d\phi\frac{1+f^{'2}(\phi)}{\left[1+f^2(\phi)+f^{'2}(\phi)\right]^{1/2}}\right]^{-1} \!\!\!\! \quad. 
\label{A}
\end{eqnarray}
Here, primes denote derivatives with respect to $\phi$. Adding the flat space energy $E_f=\pi Kq^2\ln\frac{R}{a_0}$ to the self energy, $E_s= -\pi Kq^2V = 2\pi^2Kq^2As\ln\frac{R}{a_0}$,  we obtain the total vortex energy for a singular distribution of Gaussian curvature 
\begin{eqnarray}
E_v = \pi Kq^2(1+2\pi As)\ln\left(\frac{R}{a_0}\right) \!\!\!\! \quad .
\label{cone2}
\end{eqnarray}
As a check, upon evaluating Eq. (\ref{cone2}) for a simple cone (positive delta Gaussian curvature) described by $f(\phi)=m$, $A = \frac{\sqrt{1+m^2}}{2\pi}$ and $s=1-\frac{1}{\sqrt{1+m^2}}$, we recover the result obtained in Eq. (\ref{cone1}).
\\
\\
\noindent Next, consider a saddle: a surface with negative delta Gaussian curvature. We take the simplest surface described by a height function, $h(\rho,\phi)=m\rho\cos(2\phi)$. The derivation for the total vortex energy proceeds exactly as outlined above, substituting for $f(\phi)=\text{cos}(2\phi)$ into the general expressions obtained in Eqs. (\ref{spm}-\ref{cone2}). Further, the membrane bending energy can be evaluated from Eq.\ (\ref{Hamiltonian1}) where, for our assumed height function, the mean curvature takes the form \cite{Deem:1996p15},
\begin{eqnarray}
M  = \frac{\left[1+f^2(\phi)\right]\left[f(\phi)+f''(\phi)\right]}{2\rho\left[1+f^2(\phi)+f'^2(\phi)\right]^{3/2}} \label{MeanCurv} \!\!\!\!\!\! \quad .
\end{eqnarray}
\\
\noindent Since this surface is not azimuthally symmetric, the resultant expressions can only be expressed as integrals over $\phi$. However, restricting ourselves to small deviation from flatness, i.e., $m<1$, we can expand $H_{\kappa}$, $A$ and $s$ in a perturbation series in $m$ and integrate the resulting expressions to obtain the following form, correct to order $\cal{O}$$(m^2)$ for the combined bending and vortex energies
\begin{eqnarray}
\begin{split}
E  = \pi Kq^2\left[1 + \left(\frac{9}{2}r-\frac{3}{4}\right)m^2 + {\cal{O}}(m^4) \right]\ln\left(\frac{R}{a_0}\right) \!\!\!\!\!\! \quad .\label{eqn_mexpansion}
\end{split}
\end{eqnarray}
Here, we have introduced a dimensional parameter, $r=\frac{\kappa}{K}$ that serves to quantify the competition between condensate kinetic energy and membrane bending energy.
\\
\\
\noindent Inspection of Eq. (\ref{eqn_mexpansion}) reveals a critical $r_c \sim \frac{1}{6}$, below which the total energy of the buckled membrane ($m\ne0$) is less than its flat counterpart ($m=0$). Further, due to the quadratic dependence on defect charge $q$, this result is independent of the sign of the vortex defect. Hence both, vortices and anti-vortices will buckle the membrane into a saddle shape. 
\begin{figure}
\includegraphics[width=75mm]{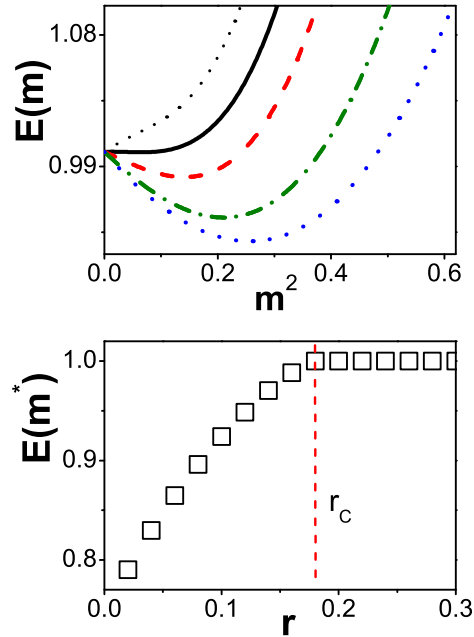}
\caption{\label{fig:fig2} (Color online) Normalized energy $E(m) = E/[(\pi K\ln(\frac{R}{a_0})]$ as a function of parameter $m^2$ for $r=\frac{4}{3} r_c$ (small dots black), $r=r_c$ (solid line black),$ r=\frac{3}{4}r_c$ (solid red), $r = \frac{1}{2} r_c$ (dash-dot green) and $r = \frac{2}{5} r_c$ (dot blue), where $r_c \sim 0.17$ is the critical ratio of membrane bending rigidity $\kappa$ to superfluid stiffness $K$, below which the membrane can buckle. For $r<r_c$, the normalized energy curves show a minima below 1(top). Bottom plot shows $E(m^*)$ as a function of $r$ also for the saddle surface, where $m^*$ is the value of $m$ for which $E(m)$ has a minimum. The plateau for $r>r_c=0.17$ shows that the membrane cannot buckle for a ratio higher than the critical value.}
\end{figure} 
\\
\\
\noindent In order to determine the shape of the buckled membrane, we expand Eq. (\ref{eqn_mexpansion}) to order $\cal{O}$$(m^6)$ and evaluate $m^*$ corresponding to a minimum in the total energy. In Fig.\ \ref{fig:fig2}, we plot the total energy $E$ so obtained, normalized by $\pi K\ln\left(\frac{R}{a_0}\right)$, against the parameter $m$ for different choices of $r$. As the ratio is decreased below the critical value $r_c = 0.17$, the energy (corresponding to red, green and blue curves) has a minimum at $m^*$ indicating that, buckling of the underlying membrane into a saddle shape is energetically favourable. 
\\
\\
\noindent We next test the validity of our ansatz for the shape of the buckled membrane by expressing the height function $h(\rho,\phi)=\rho\sum _i \left[a_i\cos(2i\phi) + b_i\sin(2i\phi)\right]$, in the form of a truncated Fourier series. Thus, we numerically seek the coefficients $\{a_i,b_i\}$ that minimizes the total Hamiltonians appearing in Eqs. (\ref{Hamiltonian1}-\ref{Hamiltonian2}), by approximating integrals with finite sums. Note, the height function used in the perturbation expansion method corresponds to retaining just the first term $a_1$ in this series. The lower panel in Fig.\ \ref{fig:fig2}, shows the energy so obtained as a function of $r$, where we have used the notation $E(m^*)$ to denote the energy obtained by solving for the Fourier coefficients that minimize the Hamiltonians even if we are using more than one term in the Fourier series expansion. Above $r_c$, the total energy remains constant, indicating that for a large ratio there is no choice of coefficients that minimizes the total energy and, therefore, membrane buckling is not favourable. However, below $r_c$, the total energy decreases as $r$ is reduced, thus confirming for both a vortex and an anti-vortex, the buckling of the underlying membrane into a saddle with the dominant contribution to the energy coming from only the first term in the Fourier series, in agreement with our analytical ansatz. 
\\
\\
\noindent Although classical screening of defects by curvature of matching sign is absent for the condensate, inspection of Eq. (\ref{eqn_mexpansion}) suggests that the reduced vortex energy, $E$, (dressed by bending contributions) can lower the critical temperature, $T_c$, for Kosterlitz-Thouless transition on a flexible substrate. This is apparent from estimating $T_c$ by balancing the vortex energy with its entropy, $S$,  \cite{Kosterlitz:1973p50}. For a two dimensional membrane of size $R$, the number of possible positions of a vortex is still proportional to $(\frac{R}{a_0})^2$ (as in flat space). Any geometric corrections are sub-leading in the limit of a large system size and one recovers the familiar result $S \approx k_{B} \ln(\frac{R}{a_0})^2$, where $k_B$ is the Boltzmann constant. By contrast, deformation of the underlying metric ($m \ne 0$) changes the pre-factor of the logarithmic divergence of the energy $E$ in Eq. (\ref{eqn_mexpansion}). At the transition, the free energy $F = E - T_c S = 0$ vanishes, leading to a reduced critical temperature
\begin{eqnarray}
T_c = \frac{E}{2k_{B}\ln\left(\frac{R}{a_0}\right)}.
\end{eqnarray}
For instance, if the bending rigidity is chosen to be below the critical value so that $r=\frac{1}{2}r_c$, we can read-off from the inset of Fig. (\ref{fig:fig2}) the corresponding shape of the saddle $m^2\sim0.21$ yielding, $T_c \sim 0.475\pi K$, whereas for a flat membrane with $m=0$, the corresponding temperature will be $T_c \sim 0.50\pi K$, with $K$ of the order of 40 Kelvin for a superfluid film of thickness around 100 Angstroms.
\\
\\
\noindent To conclude, we have demonstrated a striking manifestation of the effect of collective excitations (arising from a conformal anomaly in the kinetic energy functional of the condensate) on the mechanical stability of a membrane supporting the condensate. Unlike the prototypical buckling instability resulting from the screening of a classical defects charge by curvature, quantum buckling is a global feature, forced by an energetically more favourable target metric. The symmetry of the buckled shape with respect to defect charge, corroborates a lack of curvature-defect charge screening mechanism and is a consequence of the absence of a direct coupling between scalar quantum order parameter and geometry. In addition, we find that mechanical buckling lowers the temperature of the KT transition, providing us with an additional experimental observable for quantum buckling. Possibilities for future work could include, studying the effect of multiple interacting defects and the resultant formation of ripples on free standing membranes \cite{Guinea:2008,Katnelson:2007p858}. 
\\
\\
\noindent {\it Acknowledgments} We thank A. Turner, A. Achucarro, A. Boyarsky, L. Rademaker, L. Gomez and W.T.M. Irvine for helpful discussions.  NU gratefully acknowledges financial support from the Stichting voor Fundamenteel Onderzoek der Materie (FOM). 


\end{document}